\documentclass[%
 reprint,
 amsmath,amssymb,
 aps,
]{revtex4-2}

\usepackage{graphicx}% Include figure files\

\usepackage{dcolumn}% Align table columns on decimal point
\usepackage{bm}% bold math
\usepackage{float}
\usepackage{physics}
\begin{document}

\preprint{APS/123-QED}

\title{Anderson localization versus hopping asymmetry in a disordered lattice}% Force line breaks with \\

\author{E. T. Kokkinakis}

\author{K. G. Makris}%

\author{E. N. Economou}
\affiliation{%
 Physics Department, University of Crete, 71003, Heraklion, Greece
}%
\affiliation{%
 Institute of Electronic Structure and Laser (IESL), FORTH, 71110 Heraklion, Greece
}%
\date{\today}% It is always \today, today,

\begin{abstract}
In the framework of non-Hermitian photonics, we investigate the interplay between disorder and non-Hermiticity in a one-dimensional Hatano-Nelson lattice. While Anderson localization dictates the wave's evolution in conservative random systems, the introduction of non-Hermiticity tends to force the beam to unidirectionally propagate towards one edge of the potential due to the existence of skin modes. As we show, the antagonism between these effects results in qualitatively distinct phases of wave diffraction, including counter-intuitive characteristics regarding the relationship between the strength of disorder and the wavepacket's velocity.
\end{abstract}

\maketitle

%\tableofcontents

\section{INTRODUCTION}
The concept of Anderson localization, which predicts the possibility of suppressed diffusion for waves traveling in disordered media, has been pivotal in solid state physics research for more than sixty years \cite{anderson}. This phenomenon has been extensively studied both theoretically and experimentally in the realms of quantum and classical wave physics \cite{Akkermans, anderson_2}.
Its significance is evident in its impact across other fields beyond condensed matter physics, where direct experiments are not hindered by many-body interactions and temperature-dependent effects. Those areas include disordered photonics and imaging \cite{segev_2013, wiersma_2013}, Bose-Einstein condensates \cite{damski_2003}, and acoustics \cite{condat_1987}. However, aside from a few exceptions, such as random lasers \cite{wiersma_2008,cao,vanneste}, the research in wave localization has been dedicated to hermitian systems where power is conserved.  

Over the last decade, intense research activity has been conducted on non-Hermitian Hamiltonians, particularly in the fields of optics and photonics \cite{christodoulides_2003}, where the currently existing experimental methods \cite{wiersma_1997, lagendijk,schwartz_2007,lahini_2008} have led to the emergence of non-Hermitian photonics as a new frontier of optics \cite{el_ganainy_2018,makris_2008,Musslimani_2008, makris_2010, ruter_2010, regensburger_2012, feng_2013, zhang_2018, hodaei_2014} and the most relevant and successful platform of non-Hermitian physics. The main avenue of realizing non-Hermitian Hamiltonians, is by designing structures with spatially distributed loss and/or gain materials. Unlike solid state or quantum set ups, such dissipative materials due to dispersion and amplification due to lasing, are readily available in semiclassical optics.  

Nevertheless, non-Hermiticity can also be theoretically attained by implementing asymmetric couplings to a superconducting system through an imaginary-gauge field, as suggested by Hatano and Nelson in 1996 \cite{Hatano_1996}. Such an asymmetry, has as a direct outcome the propagation of any initial excitation towards one preferred direction of the potential \cite{Hatano_1997, longhi_2015}. Their work demonstrated that asymmetric nearest-neighbor hopping terms in a one-dimensional disordered tight binding lattice induce delocalization. The difficulty of physically realizing such an imaginary gauge field, made the Hatano-Nelson model a toy model for complex localization studies on mathematical physics \cite{feinberg_1997}. However, over the last few years there has been a keen interest regarding its experimental realization in optics \cite{liu_2021, liu_2022, gao_2023}. An intriguing feature of the Hatano-Nelson model is the strong dependence of its spectral characteristics on the boundary conditions. Under periodic boundary conditions (PBC), its eigenvalue spectrum forms an ellipse in the complex plane, where the eigenmodes are extended Bloch modes \cite{Hatano_1997}. Conversely, under Dirichlet or more commonly called open boundary conditions (OBC), the eigenvalues shrink on the real axis in the interior of the ellipse, whereas the eigenmodes get exponentially localized to one end of the lattice, showing the well-studied non-Hermitian Skin Effect (NHSE) \cite{lee_2016, yao_2018, gong_2018}. This phenomenon has drawn significant attention within the context of topological photonics \cite{harari_2018, bandres_2018, liu_2021}, which studies topologically nontrivial states \cite{okuma_2020}, and has led to counter-intuitive theoretical results with respect to bulk–boundary correspondence,
thereby leading to new definitions of topological invariants
for non-Hermitian systems \cite{li_2020,longhi_2021_1, dobrykh_2018,hang_2021,ezawa_2022,faugno_2022}.  Experimentally, numerous
novel applications have been demonstrated, such as topological lasers \cite{liu_2022,zhu_2022,harari_2018,bandres_2018} and funneling of light \cite{weidemann_2020}. Beyond topological photonics, recent theoretical studies explore the interplay between NHSE and nonlinear optical phenomena \cite{Komis_23}.

Recently, there has been a renewed interest in non-Hermitian disordered problems \cite{makris_2017, makris_2020,tzortzakakis_2020,huang_2020,tzortzakakis_2020_2}, stemming from the synergy of Anderson localization and the spatial engineering of gain/loss profiles. This new pespective on non-Hermitian Anderson localization \cite{Sukhachov_2020,liu_2020,kawabata_2021, liu_2020_b, leventis_2022}, wherein precisely controlled experiments can be performed, provides a novel and largely unexplored territory of localization in complex media. Two representative examples of intricate wave dynamics are the constant-intensity waves despite strong localization \cite{makris_2017, makris_2020} and sudden Anderson jumps \cite{leventis_2022, tzortzakakis_2020_2, weidemann_2021_} in uncorrelated disordered lattices. Such ideas have been demonstrated experimentally in acoustics \cite{rivet_2018} and optics \cite{steinfurth_2022}, paving thus the way for new aspects of Anderson localization in open systems.  

In this context, our paper's objective is to investigate the antagonism between uncorrelated on-site disorder and non-Hermiticity in a Hatano-Nelson lattice under OBC and the NHSE. In particular, we are interested on understanding the "long-term" limit, i.e., at large propagation distances, of a diffracted wavepacket, in such type of random lattices. While some related studies have examined the dynamics of non-Hermitian systems \cite{longhi_2015, longhi_2022, longhi_2021,orito_2022}, to our knowledge, there has been no systematic study of the asymptotic behavior of waves in systems exhibiting both disorder and asymmetric couplings.  In simpler terms, given a certain level of disorder and coupling asymmetry, whether the wave remains localized at its initial position, or it asymmetrically propagates towards the edge of the lattice is the central question of our work. We are thus delineating the boundaries between the previously described "localized" and "delocalized" phases and an emerging "intermediate" phase.  Furthermore, we are provide qualitative explanation of these three phases based on spectral study on the one hand, and quantitative analysis based on statistical averaging on the other hand.  The relation of the wavepacket's velocity to disorder strength is also discussed. 

%%%%%%%%%%%%%%%%%%%%%%%%%%%%%%%%%%%%%%%%%%%
%%%%%%%%%%%%%%%%%%%%%%%%%%%%%%%%%%%%%%%%%%%
\section{MODEL AND PERTINENT EXAMPLES OF DISORDER}

\subsection{The disordered Hatano-Nelson Model}

Our study begins with a one-dimensional (1D) disordered Hatano-Nelson lattice consisting of $N$-coupled waveguides (indexed by $n\in	\{ 1,2,...,N \}$) that are characterized by asymmetric couplings. Within the paraxial approximation, the evolution of waves in this lattice is described by the coupled-mode equations \cite{christodoulides_2003}:
\begin{equation}
\label{hatano_nelson}
i\frac{\partial \psi_{n}}{\partial z}
+ e^{-h}\psi_{n+1}+e^{h}\psi_{n-1}+\epsilon_{n}\psi_{n} = 0
\end{equation}
where $h \in \mathbb{R}^{+}$ represents the asymmetric hopping parameter that introduces non-Hermiticity, and $\psi_{n}$ denotes the complex amplitude of the field's envelope at position $n$ at propagation distance $z$. The on-site potential, $\epsilon_{n}$, takes values from a rectangular distribution within the interval $[-W/2, W/2]$, where $W\in \mathbb{R}^{+}$ is the disorder strength. 
\begin{figure}[!htbp]
	\includegraphics[scale=0.47]{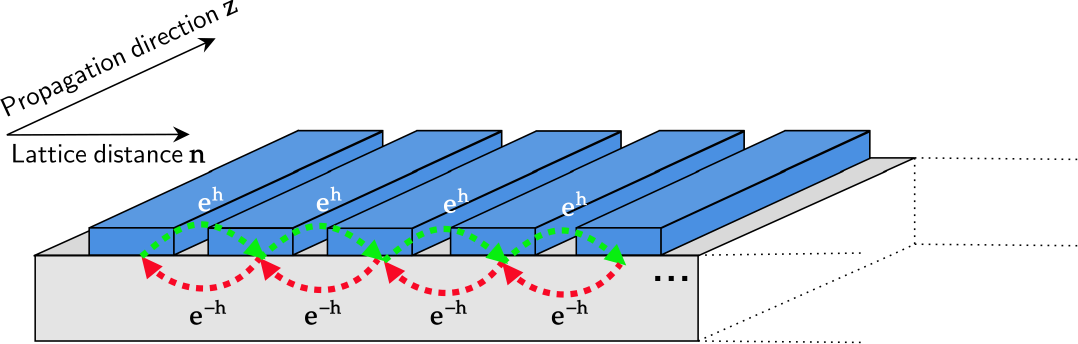}% Here is how to import EPS art
	\label{schematic}
	\caption{Schematic representation of the waveguide lattice examined in this paper. The channels are coupled asymmetrically, resulting in a preferential direction for light propagation (from left to right for $h>0$). Notably, the propagation distance $z$ serves as the optical analogue of time $t$ in Schrödinger's equation, due to the mathematical equivalence of the latter with the discrete paraxial Eq.(\ref{hatano_nelson}).}
\end{figure}

Thus, our system is governed by a real-element non-Hermitian  Hamiltonian-$\hat{H}$ with random diagonal elements. The eigensolutions of the right eigenvalue problem associated with this Hamiltonian are expressed as $\ket{\psi_{j}(z)}=\ket{u_{j}^{R}}e^{i\omega_{j}z}$, where $\ket{u_{j}^{R}}$ is one right eigenvector of the set $\{\ket{u_{k}^{R}}\}$, and $\omega_{j}$ is the corresponding eigenvalue, i.e:
\begin{equation}
	\label{right_ham}
	\hat{H}\ket{u_{j}^{R}}=\omega_{j}\ket{u_{j}^{R}}
\end{equation}
Regarding the associated boundary conditions, we consider the waveguide channels to form a linear chain, i.e. OBC ($\psi_{0}=\psi_{N}=0$) hold, so that the non-degenerate spectrum of $\hat{H}$ is entirely real.
Since the Hamiltonian associated with  Eq.(\ref{hatano_nelson}) is real, meaning that $H^{\dagger}=H^{T}$ and also $\omega_{j}^{*}=\omega_{j}$, the left eigenvalue problem associated with Eq.(\ref{right_ham}) reads as
\begin{equation}
	\label{left_ham}
	\hat{H}^{T}\ket{u_{j}^{L}}=\omega_{j}\ket{u_{j}^{L}}
\end{equation}
The right and left eigenvectors of the Hamiltonian satisfy the biorthogonality condition
\begin{equation}
	\label{bio}
	\bra{u_{i}^{L}}\ket{u_{j}^{R}}=\delta_{ij}
\end{equation}
It is noteworthy that by applying an imaginary gauge transformation $\alpha_{n} \equiv \psi_{n}e^{-hn}$ to Eq. (\ref{hatano_nelson}), one can derive the couple-mode theory equation for the symmetric hermitian Anderson model:
\begin{equation}
\label{anderson}
i\frac{\partial \alpha_{n}}{\partial z}
+ \alpha_{n+1}+\alpha_{n-1}+\epsilon_{n}\alpha_{n} = 0.
\end{equation}
The amplitude of the field's envelope on lattice site $n$, denoted as $\psi_{n}(z)\equiv\bra{n}\ket{\psi}$, can be expressed in the basis of the Hamiltonian's right eigenvectors $\ket{u_{j}^{R}}$ as follows:
\begin{equation}
    \label{expansion}
    \psi_{n}(z)=\sum_{j=1}^{N}   C_{j}u_{j,n}
\end{equation}
where $u_{j,n}\equiv  \bra{n}\ket{u_{j}^{R}}$ and $C_{j}\equiv \bra{u_{j}^{L}}\ket{\psi}$.
Since the system under discussion exhibits an entirely real spectrum, the projection coefficients $C_{j}$ evolve as:
\begin{equation}
C_{j}=\bra{u_{j}^{L}}\ket{\psi_{0}}e^{i\omega_{j}z}\equiv C_{j,0}e^{i\omega_{j}z}
\end{equation}
meaning that their magnitudes are independent of the propagation distance $z$, and depend only on the initial condition. In other words, the propagation dynamics is an effect of interference between the different eigenmodes and does not result from different rates of amplification or dissipation of some modes \cite{leventis_2022}.

However, the system is not conservative since the Hamiltonian is non-Hermitian.  Therefore its optical power 
\begin{equation}
    \mathcal{P}(z)\equiv\bra{\psi}\ket{\psi}=\sum_{n=1}^{N} |\psi_n(z)|^2
\end{equation}
varies during propagation. Such an effect- power oscillations- is typical in many non-Hermitian systems \cite{makris_20086}. Thus, a normalized amplitude for the field's envelope can be introduced at every propagation distance $z$, namely
\begin{equation}
    \label{normalization}
    \phi_{n}\equiv \psi_{n}/\sqrt{\mathcal{P}(z)}
\end{equation}
in order to better capture the wavepacket's mean position. This renormalization does not affect the spatial distribution of the wavefunction on each propagation step and is experimentally feasible \cite{weidemann_2021_}.  
However, in contrast to cases where the on-site potential $\epsilon_{n}$ has an imaginary part, reflecting the presence of gain and loss in the system, and where the power increases exponentially as the most gainy eigenstates dominate \cite{leventis_2022}; in our case, the power oscillations are bounded. This can be directly shown, based on the reality of the associated eigenspectrum. In particular, the optical power $\mathcal{P}(z)$ can be expressed as
\begin{equation}
\mathcal{P}(z)=
\sum_{i=1,j=1}^{N,N}
(C_{j}\ket{u_j^{R}})(C_{i}\ket{u_i^{R}})^{*}=
\sum_{i=1,j=1}^{N,N}
\Delta_{i,j}e^{i(\omega_{j}-\omega_{i})z}
\end{equation}
where
\begin{equation}
    \Delta_{i,j}\equiv\bra{u_{i}^{R}}\ket{u_{j}^{R}}\bra{u_{j}^{L}}\ket{\psi_{0}}\bra{\psi_{0}}\ket{u_{i}^{L}}  
\end{equation}
After some straightforward manipulation and using the symmetry of index interchange  $\Delta_{i,j}=\Delta_{j,i}$ one obtains:
\begin{equation}
    \mathcal{P}(z)=
    \sum_{i=1}^{N} 
    \Delta_{i,i}+\sum_{i=1,j=1:j\neq i}^{N,N} 
    \Delta_{i,j}\cos[(\omega_{j}-\omega_{i})z]
\end{equation}
From this relation, it is evident that
\begin{equation}
    \mathcal{P}(z)\le
    \sum_{i=1}^{N} \Delta_{i,i}+\sum_{i=1,j=1:j\neq i}^{N,N} 
    |\Delta_{i,j}|
\end{equation}
so the optical power is indeed a bounded quantity.

Some key physical quantities concerning the dynamic characteristics of the field are, the mean position and its uncertainty, that are typically defined through the following relations:

\begin{equation}
	\langle x \rangle_{z} \equiv \sum_{n=1}^{N} n \left| \phi_{n}(z) \right|^2
\end{equation}

\begin{equation}
	| \Delta x |_{z} \equiv \sqrt{\langle x^2 \rangle_{z} - \langle x \rangle_{z}^2}
\end{equation}
where
\begin{equation}
	\langle x^2 \rangle_{z} \equiv \sum_{n=1}^{N} n^{2} \left| \phi_{n}(z) \right|^2
\end{equation}

In the first part of this work, we focus on numerically simulating the propagation of normalized field envelope, $\phi_{n}(z)$, in order to understand how disorder competes with the propagation due to asymmetric couplings. In particular, we investigate the impact of the non-Hermiticity parameter $h$ in lattices with different values of disorder strength $W$, under single- channel excitation at one end of the lattice, i.e.
\begin{equation}
    \label{initial}
    \psi_{n}(z=0)=\delta_{n,1}
\end{equation}
In the following paragraphs, we examine pertinent examples of diffraction in a random Hatano-Nelson lattice for a single realization at different disorder strengths: low (B), intermediate (C), and high (D).

\subsection{Low disorder strength $W=1$}
Here we examine the case of weak disorder ($W=1$) for a single realization. The single-channel initial condition excites a superposition of many eigenstates, which leads to the spreading of the wavefunction across a bounded spatial extent, as we can clearly see in Fig. 2(a) for $h=0$ (hermitian Anderson lattice).

As we increase the non-Hermiticity parameter $h$, the wave undergoes partial delocalization towards the preferred direction of the lattice, as illustrated in Fig. 2(b). In this case, the position uncertainty $|\Delta x| _{z}$ (Fig. 2(d)) of the wave remarkably increases, since it is extended across the entire lattice.  However, regardless of the value of the propagation distance $z$, the wave does not get entirely displaced towards the end of the lattice. 

At this point, let us note the following: for a lattice of $N$ sites, it is considered sufficient to draw conclusions about the long-term asymptotic behavior by examining its behavior up to a propagation length a few times greater than the lattice size $N$, for example, $z_{\text{max}} = 10N$. For propagation distances  significantly larger than the lattice size, no significant deviations were observed.

\begin{figure}[!htbp]
\includegraphics[scale=0.41]{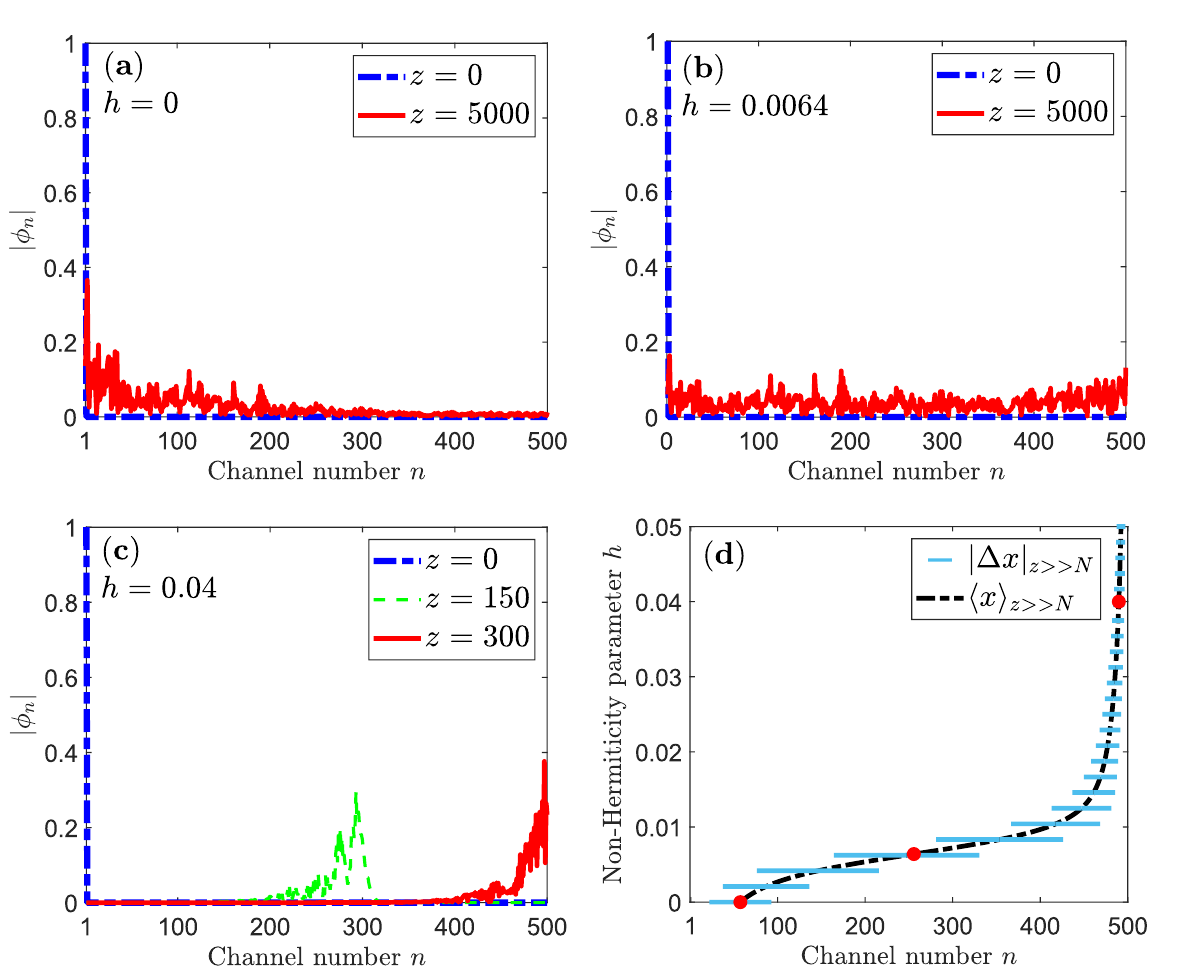}
\label{single_w_1}
\caption{Single realization of weak disorder (\(W=1\)). Characteristic cases of the propagation dynamics of the wavefunction \(|\phi_{n}|\) on a Hatano-Nelson lattice with \(N=500\). (a) For \(h=0\), the wave remains localized close to the excitation edge (b) For \(h=0.0064\), the wave becomes partially delocalized, spreading across the entire lattice, increasing its mean position and  uncertainty. (c) For \(h=0.04\), the wave fully delocalizes from its initial position and reaches the end of the lattice at a propagation distance \(z < N\). (d) The mean position, \(\langle x \rangle_{z \gg N}\), and its uncertainty, \(|\Delta x|_{z \gg N}\), for a propagation distance \(z=5000 \gg N\) versus the non-Hermiticity parameter \(h\). The three red dots correspond to the cases presented in (a), (b) and (c), respectively.}
\end{figure}

For higher values of \(h\), the wave undergoes complete delocalization with respect to its initial position, and reaches the end of the lattice at a propagation distance smaller than the lattice size $N$ (Fig. 2(c)), manifesting the non-Hermitian skin effect. Such displacement occurs in an almost ballistic fashion (Fig. 3(a)), i.e. the wave's velocity, defined as \(u \equiv d\langle x \rangle/dz\), remains nearly constant, after some initial acceleration \cite{longhi_2022} (Fig. 3(b)). In this range of \(h\) values, it is evident that non-Hermiticity completely dictates the propagation over disorder. 

\begin{figure}[!htbp]
\includegraphics[scale=0.47]{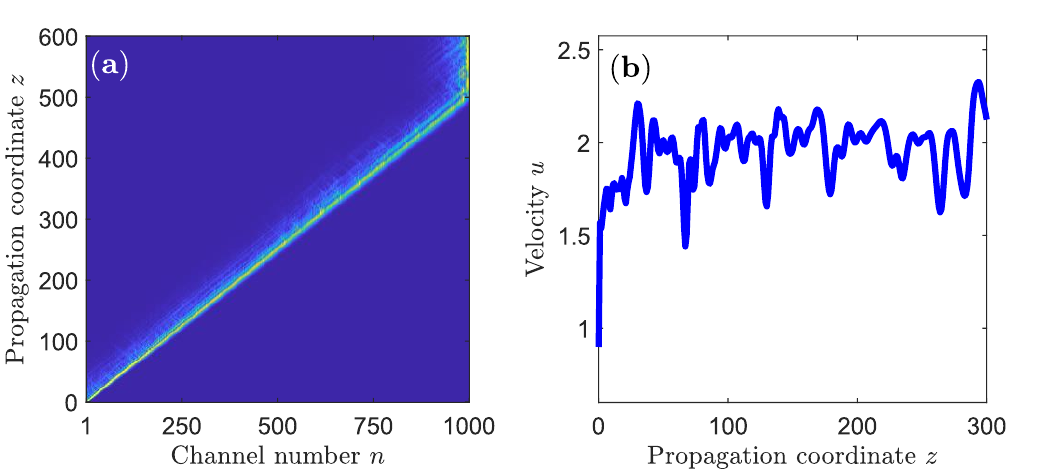}
\label{ballistic}
\caption{Demonstration of ballistic dynamics on a Hatano-Nelson lattice with $N=1000$ sites, \(W=1\), and \(h=0.04\): (a) Wavefunction \(|\phi_{n}(z)|\)   (b) The derivative of the mean position over the propagation distance, \(u \equiv d\langle x \rangle/dz\). After an initial acceleration the wave propagates with a velocity fluctuating around the value \(u \approx 2\). }
\end{figure}

\subsection{Intermediate disorder strength $W=3$}

Let us now consider a higher value of disorder strength, namely ($W = 3$). Our results are presented in Fig.4. In fact, we observe a range of non-zero values for the non-Hermiticity parameter $h$, for which the wave remains entirely localized within a narrow region close to its initial location, as demonstrated in Fig. 4(a). This localization persists regardless of the propagation distance, similar to the behavior observed in a hermitian lattice. Within this range of $h$-values, the wave's behavior at a given distance appears largely unaffected by non-Hermiticity (see Fig. 4(d)). If we further increase $h$, the wave begins to delocalize from its initial position across a limited region of the lattice (Fig. 4(b)), although not in a spreading manner. Instead, the wave exhibits greater localization in specific areas within this region.

For even larger values of $h$, the non-Hermitian nature of the system begins to dominate over disorder, akin to what is observed with smaller disorder strengths. However, in this case, the ballistic dynamics are replaced by a series of directed jumps toward the end of the lattice, as shown in Fig. 4(c).

\begin{figure}[!htbp]
\includegraphics[scale=0.41]{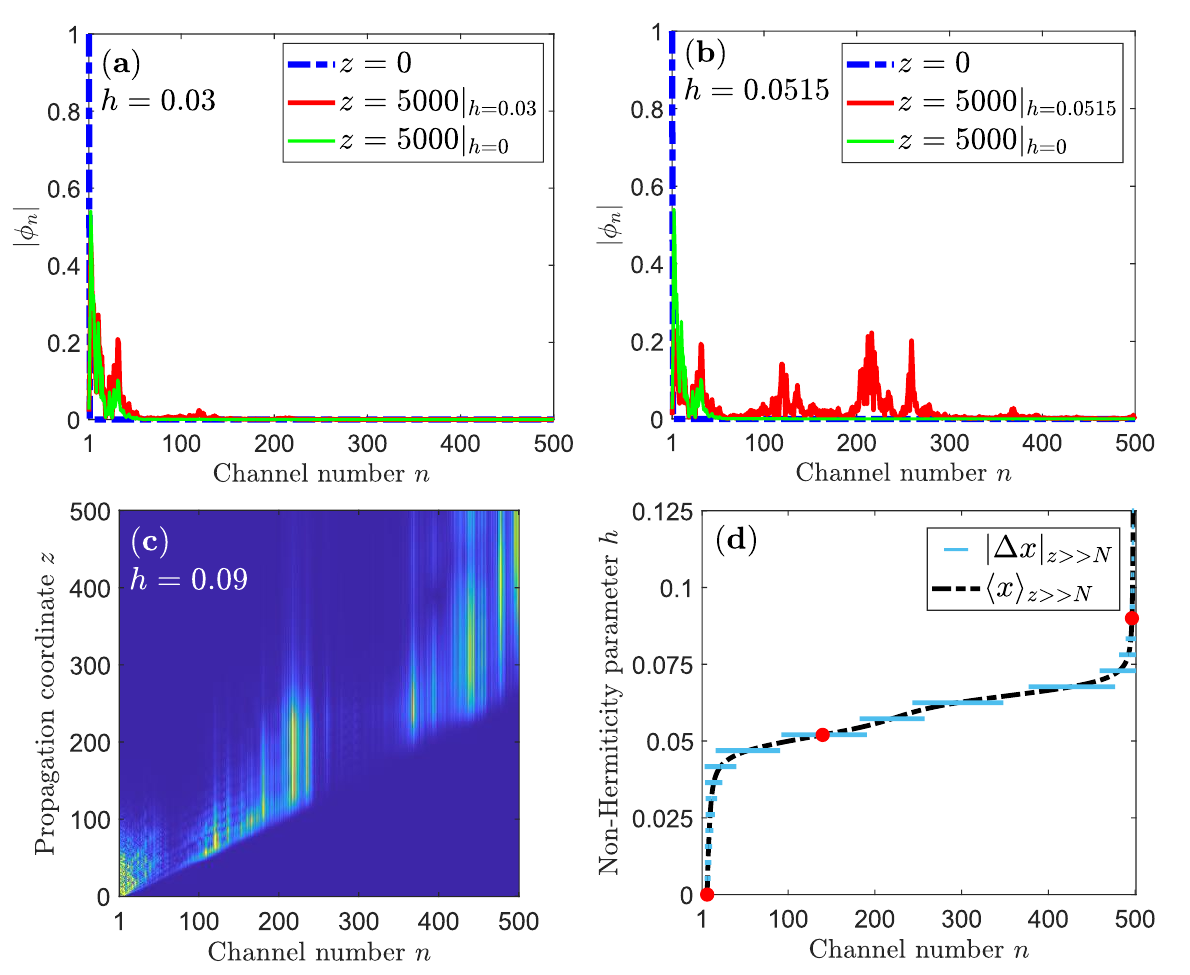}% Here is how to import EPS art
\label{single_w_3}
\caption{Single realization of intermediate disorder (\(W=3\)). Characteristic cases of the propagation dynamics of wavefunction \(|\phi_{n}|\), on a Hatano-Nelson lattice of \(N=500\) sites. (a) For \(h=0.03\), the wave remains almost completely localized to its initial location. (b) For \(h=0.0515\), the wave gets partially delocalized from its initial position (c) For \(h=0.09\), the wave gets fully delocalized from its initial position and through consecutive directional jumps shifts to the end of the lattice (d) A plot of the wave's mean position, \(\langle x \rangle_{z\gg N}\), and its uncertainty, \(|\Delta x|_{z\gg N}\), for propagation distance \(z=5000 \gg N\) as a function of $h$. The three red dots correspond to the cases presented in (a), (b) and (c), respectively.}
\end{figure}

\subsection{High disorder strength $W=6$}

We move on to the last representative example, that of high disorder, $W = 6$. In this case, the wave demonstrates resilience against delocalization from its initial position (Fig. 5(a)) over a broader range of values of the non-Hermiticity parameter, $h$, as shown in Fig. 5(d). Interestingly, in the strong disorder regime, there is a specific interval of $h$-values where the wave does not significantly diffract, but either stays localized in a different region or maintains partial localization across two or more distant regions, with optical intensity oscillating between them, as depicted in Fig.5(b). This intermediate phase is qualitatively distinct from the lower disorder regime, where spreading occurred, and its specific characteristics are highly dependent on the potential realization. For sufficiently large values of the non-Hermiticity parameter $h$, the non-Hermitian asymmetry of the system begins to dominate over disorder, similar to the previous case. In this regime, the wave progresses towards the end of the lattice through a series of directed jumps, as shown in Fig.5(c).
\begin{figure}[!htbp]
\includegraphics[scale=0.41]{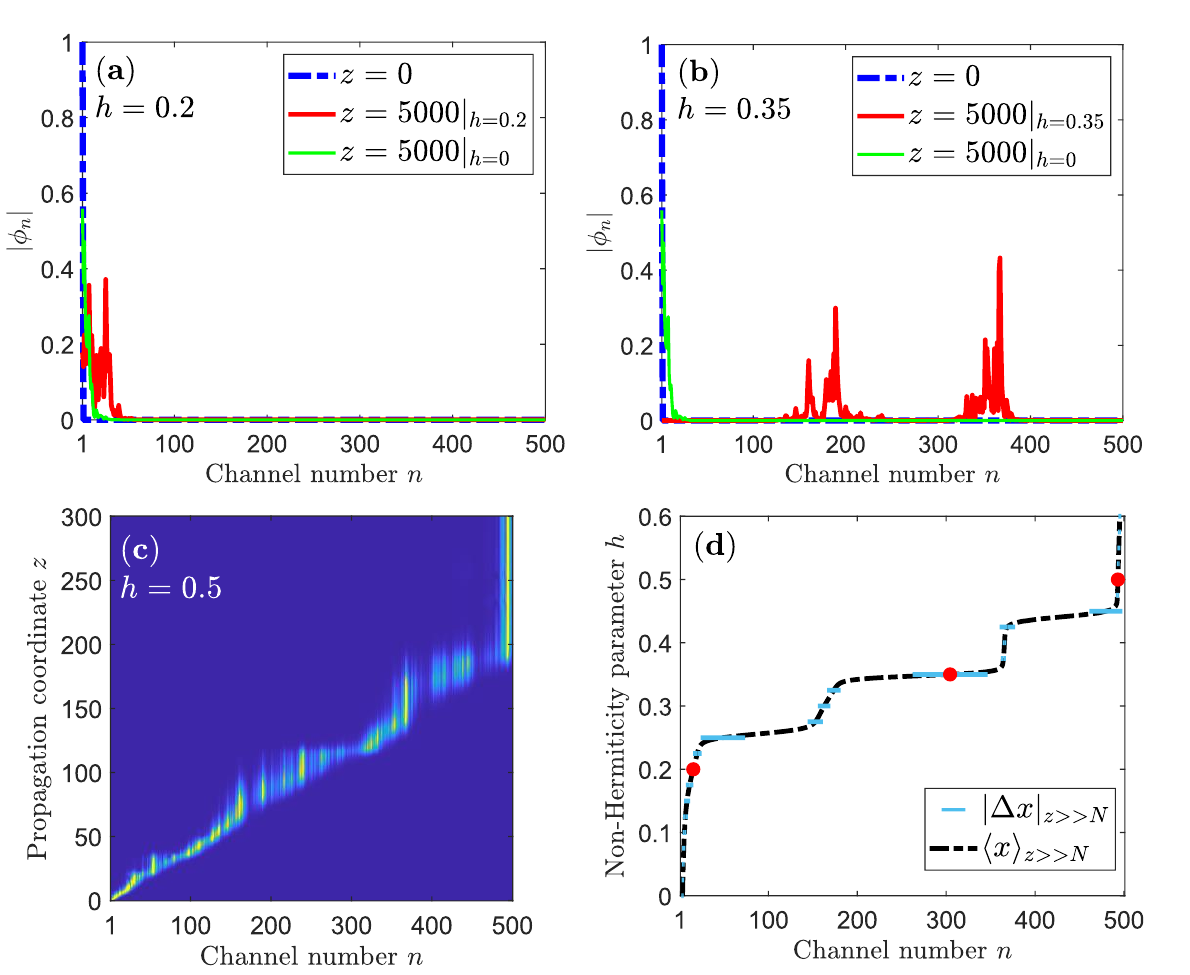}% Here is how to import EPS art
\label{single_w_6}
\caption{Single realization of strong disorder (\(W=6\)). Characteristic cases of the wavefunction's \(|\phi_{n}|\) diffraction, on a Hatano-Nelson lattice of \(N=500\) sites. (a) For \(h=0.2\), the wave remains almost completely localized to a adjacent region. (b) For \(h=0.35\), the wave moves but remains trapped between two distant regions, without ever reaching the end of the lattice. (c) For \(h=0.5\), the wave gets fully delocalized from its initial position and through consecutive directional jumps shifts to the end of the lattice (d) Wave's mean position, \(\langle x \rangle_{z\gg N}\), and its uncertainty, \(|\Delta x|_{z\gg N}\), for \(z=5000 \gg N\) as a function of $h$. The three red dots correspond to the cases presented in (a), (b) and (c), respectively.}
\end{figure}

In conclusion, the aforementioned situations correspond to three discrete phases in the propagation dynamics: localized, intermediate, and delocalized. In the low disorder regime, the first phase is not present. However, the transition zones between localized, intermediate, and delocalized phases in the wave propagation dynamics are not sharply defined for each value of the disorder strength and are highly sensitive to the specific characteristics of each potential realization. Therefore, one should conduct statistical analysis in order to draw general conclusions, as we will see in Section IV. However, before doing this, we have to systematically understand and characterize the dependence of the spatial structure of the underlying eigenmodes on both \(W\) and \(h\)-parameters in the following Section III.

%%%%%%%%%%%%%%%%%%%%%%%%%%%%%%%%%%%%%%%%%%%
%%%%%%%%%%%%%%%%%%%%%%%%%%%%%%%%%%%%%%%%%%%

\section{Localization of eigenmodes and DYNAMICS}

In this section, we are going to address the following relevant question: what enables transport for certain $W$ and $h$-values, while forcing localization for others, with respect to the three distinct phases? 
In order to answer this question, we will provide a theoretical analysis for the pertinent (single realization) cases of the previous Section II. However, we shall confine ourselves to a lattice of $N = 50$ here, as the precise computation of eigenstates of our non-Hermitian Hamiltonian for high values of $N$ is a computationally challenging issue \cite{chen_2023}. 
As mentioned in Section II, the electric field's envelope $\ket{\psi(z)}$ can be expressed using the expansion to its right eigenstates as:
\begin{equation}
    \label{expansion_2}
    \ket{\psi(z)} = \sum_{j=1}^{N} C_{j,0} e^{i\omega_{j}z}\ket{u_{j}^R}
\end{equation}
where $C_{j,0} = \bra{u_{j}^{L}}\ket{\psi_{0}}$. The propagation dynamics of a wave described by this relation is a dynamical interference between different right eigenmodes, characterized by unequal weights $C_{j,0}$ that solely depend on the initial condition and the particular realization of disorder. Interference mainly occurs between modes $\ket{u_{j}^R}$ with high absolute values of the projection coefficient $|C_{j,0}|$, since an eigenstate $\ket{u_{m}^R}$ with very small $C_{m,0}$ compared to the projection coefficient of the rest eigenstates has negligible contribution to the expansion of Eq.(\ref{expansion_2}). Note, however, that while the eigenstates are normalized to satisfy the biorthogonality condition Eq.(\ref{bio}), their normalization is arbitrary up to a multiplication factor. Therefore, it is only meaningful to compare the projection coefficients relatively to each other, rather than their absolute values. Thus we expect that when all significantly contributing eigenmodes are predominantly located on the side of the lattice near $n=1$, transport of the initial excitation towards the lattice's preferred direction becomes essentially impossible.

A useful measure for the spatial location of each right eigenstate $\ket{u_{j}^R}$  along the lattice is its center of mass $\langle n \rangle_{j}$, defined as:
\begin{equation}
    \langle n \rangle_{j} \equiv \frac{\sum_{n=1}^{N} n \abs{u_{j,n}}^2}{\sum_{n=1}^{N} \abs{u_{j,n}}^2}
\end{equation}

Here we focus on the case of strong disorder \(W=6\), and present our results in Fig.~6, Fig.~7, and Fig.~8, for three different values of the non-Hermiticity parameter \(h=0.1, 0.3, 0.55\), respectively.

In Fig.6(a), we depict the propagation dynamics of wavefunction $|\phi_{n}|$ for a single disorder realization ($W = 6$, $h = 0.1$). The wave remains localized regardless of the propagation distance, since the non-zero initial projection coefficients (Fig.6(b)) correspond to eigenstates localized near the edge $n=1$ of the lattice. Notably, for this $h$ value, the distribution of the eigenstates' centers of mass, $\langle n \rangle_{j}$, along the lattice is almost uniform, with a slightly higher concentration towards the lattice's preferred end. Nevertheless, transport is forbidden, due to negligible projection coefficients $|C_{j,0}|$ corresponding to eigenstates localized within the region $n>10$ of the lattice. 

\begin{figure}[!htbp]
\includegraphics[scale=0.47]{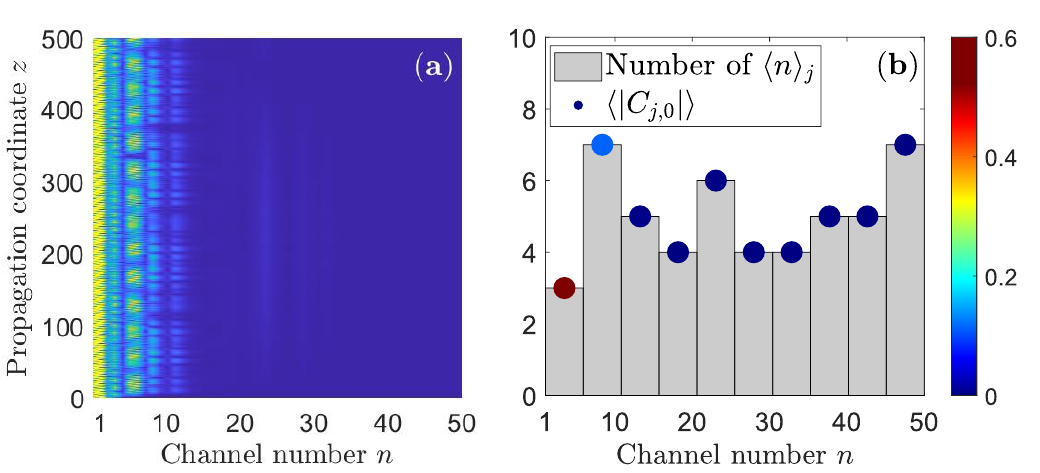}
\label{figure1}
\caption{Diffraction in a lattice of \( N = 50 \) waveguides with \( W = 6 \) and \( h = 0.1 \) (a) Magnitude of the wavefunction \( |\phi_{n}(z)| \)
(b) Spatial distribution of the centers of mass of \( N = 50 \) eigenstates along the lattice;  the height of each bar shows the number of states within each region of channels. The colored points indicate the mean magnitude of projection coefficients $\langle|C_{j,0}|\rangle$ of the eigenstates located within each bar.}
\end{figure}
For a higher value of $h = 0.3$, the system is in the intermediate phase, where the wave becomes partially delocalized, exhibiting oscillations between two different regions but failing to entirely reach the other side of the lattice, as illustrated in Fig.7(a). This could be expected based on the magnitudes of the projection coefficients (Fig.7(b)), which show their highest values for eigenstates with centers of mass located around $n=5$ and $n=30$. This is in perfect agreement to the numerically observed long-term diffraction dynamics.

\begin{figure}[!htbp]
\includegraphics[scale=0.47]{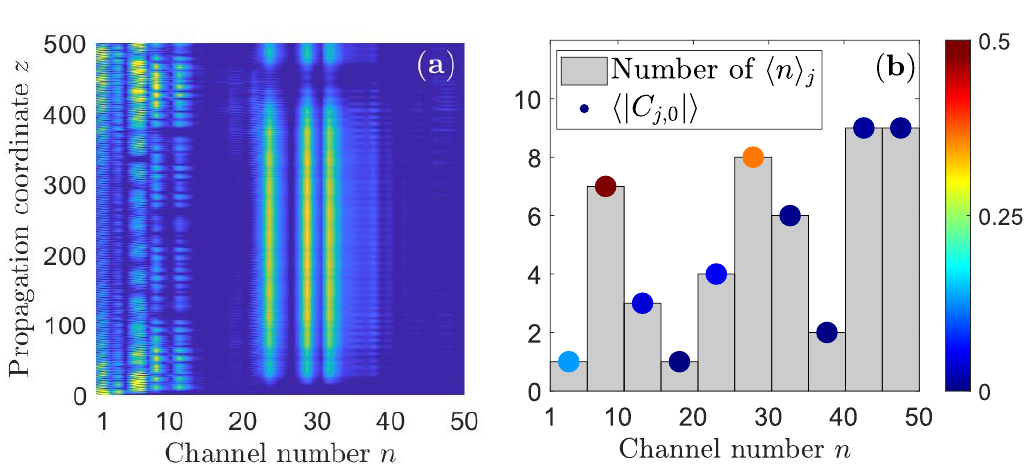}
\label{figure1}
\caption{Diffraction in a lattice of \( N = 50 \) waveguides with \( W = 6 \) and \( h = 0.3 \) (a) Magnitude of the wavefunction \( |\phi_{n}(z)| \)
	(b) Spatial distribution of the centers of mass of \( N = 50 \) eigenstates along the lattice;  the height of each bar shows the number of states within each region of channels. The colored points indicate the mean magnitude of projection coefficients $\langle|C_{j,0}|\rangle$ of the eigenstates located within each bar.}
\end{figure}

When the non-Hermiticity parameter is further increased to $h=0.55$, the system enters the total delocalization phase, as depicted in Fig.8(a). This behavior is expected, given the fact that over half of the eigenstates have shifted towards the preferred end of the lattice ($n=50$), and have the highest mean projection coefficients, as illustrated in Fig.8(b). 

\begin{figure}[!htbp]
\includegraphics[scale=0.47]{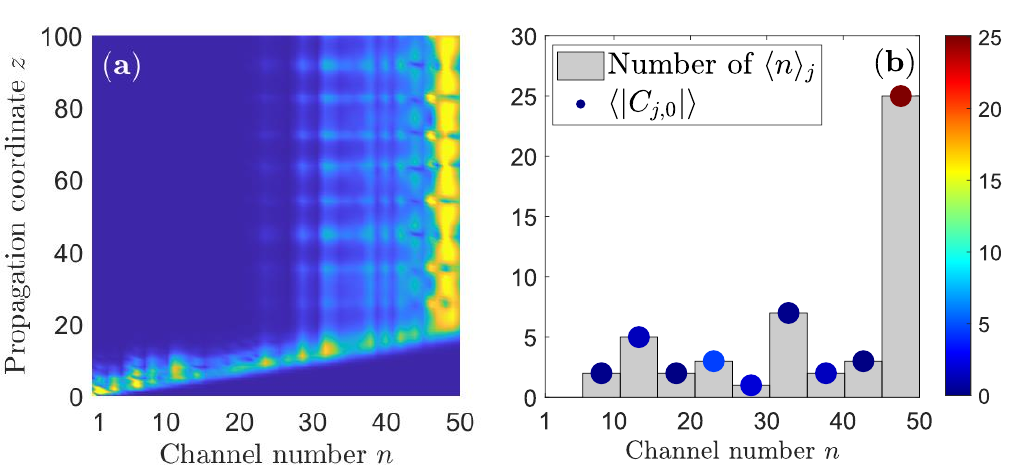}
\label{figure1}
\caption{Diffraction in a lattice of \( N = 50 \) waveguides with \( W = 6 \) and \( h = 0.55 \) (a) Magnitude of the wavefunction \( |\phi_{n}(z)| \)
	(b) Spatial distribution of the centers of mass of \( N = 50 \) eigenstates along the lattice;  the height of each bar shows the number of states within each region of channels. The colored points indicate the mean magnitude of projection coefficients $\langle|C_{j,0}|\rangle$ of the eigenstates.}
\end{figure}

At this point, we introduce another useful metric for the spatial localization degree of every eigenmode $\ket{u_{j}^R}$ which is the participation ratio $PR_{j}$, defined as:
\begin{equation}
	PR_{j} \equiv \frac{\left(\sum_{n=1}^{N} \abs{u_{j,n}}^2\right)^2}{\sum_{n=1}^{N} \abs{u_{j,n}}^4}
\end{equation}
Low values of $PR_{j}$ indicate more localized modes, and vice versa.

Now we can comment on the fact that  the dynamics depicted in Fig.8(a) ($W=6$) and Fig.3 ($W=1$) differ in their transport mechanism: directed jumps (ballistic transport) characterizing the former (latter). In both cases certain bulk eigenstates with low projection coefficients are dispersed throughout the lattice, and thus, the wave passes through them during its displacement. The qualitative difference arises from the significantly different participation ratios of bulk eigenstates. For the case of strong disorder ($W=6$) in a lattice of $N=50$ sites with $h=0.55$, the participation ratio of eigenstates with centers of mass $\langle n \rangle_{j}<40$, averaged over 50000 realizations of disorder, was calculated to be $\overline{PR_{j}}_{\text{bulk}}=1.79$. For the corresponding scenario with low disorder ($W=1$) and $h=0.1$, this quantity was determined to be $\overline{PR_{j}}_{\text{bulk}}=22.39$, reflecting the broader spatial distribution of bulk eigenstates in the lower disorder regime. In other words, we can interpret this different behavior as follows: in the high disorder regime, the bulk eigenstates are highly localized. Thus, when the wave overlaps with them along propagation, it gets confined within tight spatial regions along its path towards the lattice's end. Conversely, in the low disorder case, the bulk eigenstates exhibit a much greater spatial extent, resulting in a smoother ballistic propagation behavior.

%%%%%%%%%%%%%%%%%%%%%%%%%%%%%%%%%%%%%%%%%%%
%%%%%%%%%%%%%%%%%%%%%%%%%%%%%%%%%%%%%%%%%%%
 
\section{Statistical Analysis of Wave Dynamics}
As mentioned in Section II, the boundaries between  localized, intermediate, and delocalized phases in the wave diffraction dynamics are not sharply defined and are highly sensitive to the specific characteristics of each disorder realization. Therefore, in order to draw general conclusions, the application of statistics over many realizations of disorder is inevitable. More specifically, we considered different realizations of disorder for a lattice consisting of $N=100$ sites under various values of the disorder strength $W$ and the non-Hermiticity parameter $h$. For each realization, we calculated the mean position and position uncertainty for large distances $z=5000\gg N$. Subsequently, statistical averaging was performed over a sufficiently large number of realizations to ensure convergence of our results.

Our results are shown in Fig.9. More specifically, in Fig.9(a), the boundaries between the localized, intermediate, and delocalized regions are unveiled. The statistically averaged mean position of the wave for large propagation distances reveals that, for a significant region of the $W-h$ plane, the wave remains localized near its initial position. For each value of disorder $W$, there exists a critical value of the non-Hermiticity parameter $h_{1}$ where the system enters the intermediate phase, i.e., it becomes delocalized from its initial position but does not reach the end of the lattice; this is reflected in increased values of $\overline{\langle x \rangle _{z>>N}}$, albeit still less than $N$. Subsequently, at another critical value $h_{2}$, the system enters its delocalized phase, signified by $\overline{\langle x \rangle _{z>>N}}\sim N$. It is evident that the difference $h_{2}-h_{1}$ increases as a function of $W$. The extent of the intermediate phase is also apparent in Fig.9(b), illustrating the statistical standard deviation of $\langle x \rangle _{z>>N}$ across different realizations for varying $W$ and $h$; it remarkably increases while the system is in the intermediate phase, where it is neither entirely localized nor delocalized, and its propagation dynamics highly depends on the disorder realization. The increase of difference $h_{2}-h_{1}$ in the intermediate phase for higher values of disorder is also apparent. More over, Fig.9(c) presents the statistical average of $|\Delta x|_{z\gg N}$, that further elucidates on the boundary between localized and delocalized regions and provides valuable insight. This quantity takes significant values in the intermediate phase for low disorder since is associated with increased spreading along the lattice, whereas for stronger disorder, it decreases since position uncertainty does not arise from spreading but by the partial localization of the wave between spatially distant regions of the lattice, as discussed in Section II.

\begin{figure}[!htbp]
\includegraphics[scale=0.38]{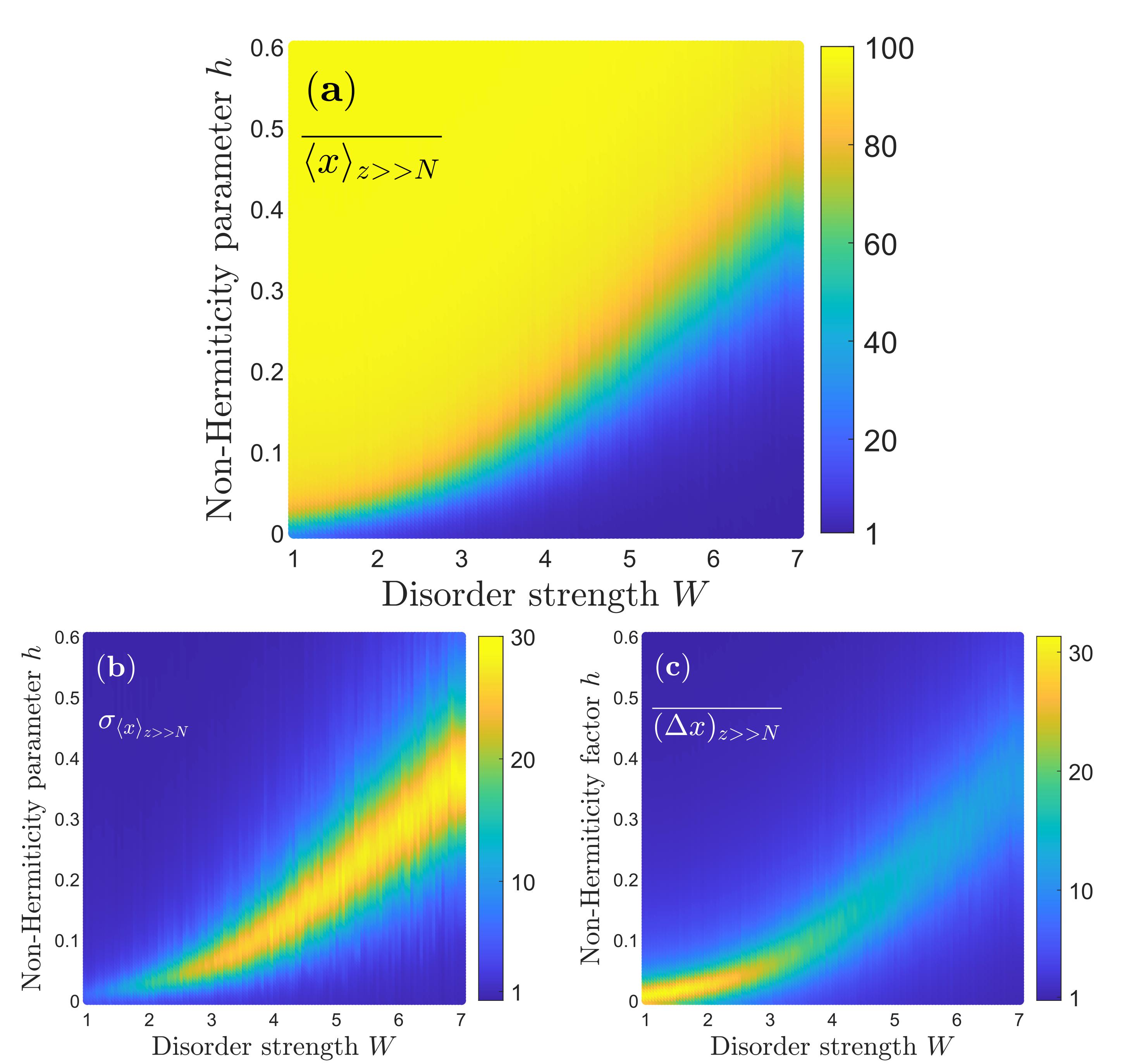}
\label{figure1}
\caption{$W-h$ planes describing the long propagation distance  ($z=5000\gg N$) dynamical behavior for a lattice of $N=100$ sites, where averaging over high enough number of realizations has been employed in order to to achieve convergence. In all three maps, the boundary between localized and delocalized phases appearing as an intermediate phase is evident. (a) Statistically averaged mean position for $z=5000$: $\overline{\langle x \rangle _{z>>N}}$ (b) Statistical standard deviation of mean position for $z=5000$: $\sigma_{\langle x \rangle _{z>>N}}$ (c) Statistically averaged position uncertainty for $z=5000$: $\overline{|\Delta x|_{z\gg N}}$.}

\end{figure}

%%%%%%%%%%%%%%%%%%%%%%%%%%%%%%%%%%%%%%%%%%%
%%%%%%%%%%%%%%%%%%%%%%%%%%%%%%%%%%%%%%%%%%%
\section{VELOCITY OF BALLISTIC PROPAGATION}
A final intriguing question that we are going to examine in this section, is whether and how the velocity of ballistic (or ballistic "on average", through directional jumps) propagation depends on the values of $h$ and $W$.

To address this question, we selected a range of values for $W$ and $h$ within the delocalized phase shown in Fig. 9(a). For these values of $W$ and $h$, 2000 realizations of potential were performed for a lattice of size $N=100$, and the mean position $\langle x \rangle$ over $z$ was calculated. Averaging over the realizations was performed, and the following quantity was obtained:
\begin{equation}
    u_{\text{avg}} \equiv \left(\frac{d\overline{\langle x \rangle}}{dz}\right)_{\delta \zeta} 
\end{equation}
where $\delta \zeta$ is the propagation distance interval within which, for each $W$ and $h$, the quantity $d\overline{\langle x \rangle}/{dz}$ is almost constant, with a tolerance of $\delta u_{tol}=0.1$ and by $\left(\cdot\right)_{\delta \zeta}$ we denote averaging over $\delta \zeta$. 

Our results are shown in Fig.10, and led us to two major conclusions.
\begin{figure}[!htbp]
	\includegraphics[scale=0.5]{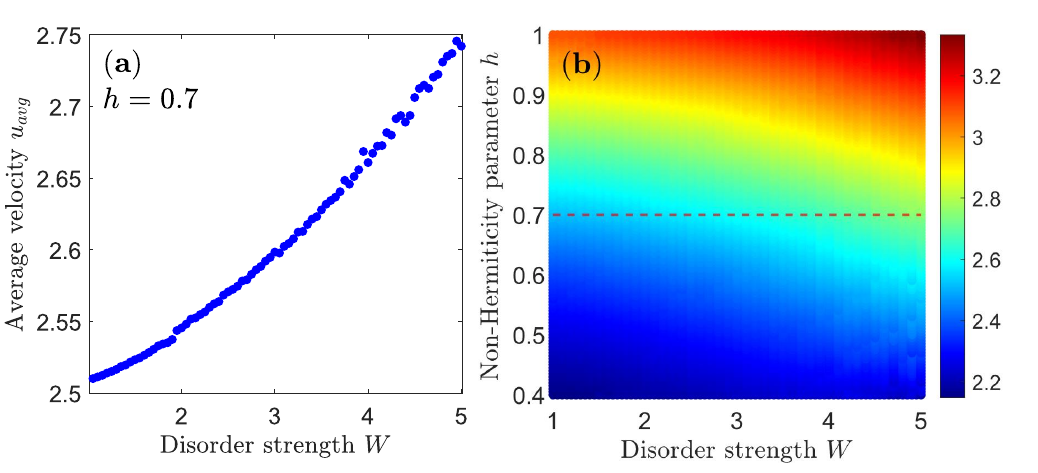}
	\centering
	\label{figure1}
	\caption{Velocity dependence on disorder and non-Hermiticity for a lattice of $N=100$ waveguides: (a) Average velocity $u_{avg}$ of propagation versus disorder strength $W$ for a fixed value $h=0.7$   (b) Average velocity $u_{\text{avg}}$ for values of $W$ and $h$ corresponding to the delocalized phase of the $W-h$ map of Fig.9. The dashed line denotes the cross section that is shown in (a).}
\end{figure}
On the one hand, for a given disorder strength $W$, the velocity increases as the non-Hermiticity parameter $h$ increases. This is due to the increased asymmetry in the hopping terms between the lattice sites, and is expected.

On the other hand, for a given non-Hermiticity parameter $h$, the average velocity increases monotonically with disorder strength $W$ (Fig. 10(b)). This observation is counter-intuitive; higher disorder implies a greater tendency for localization, which one could expect to result in lower average velocity. However, similar results of disorder-induced transport enhancement have been reported in recent studies on quasi-periodic crystals \cite{longhi_2021, orito_2022}. It appears that this effect is universal and holds in disordered non-Hermitian systems as well.
\begin{figure}[!htbp]
	\includegraphics[scale=0.4]{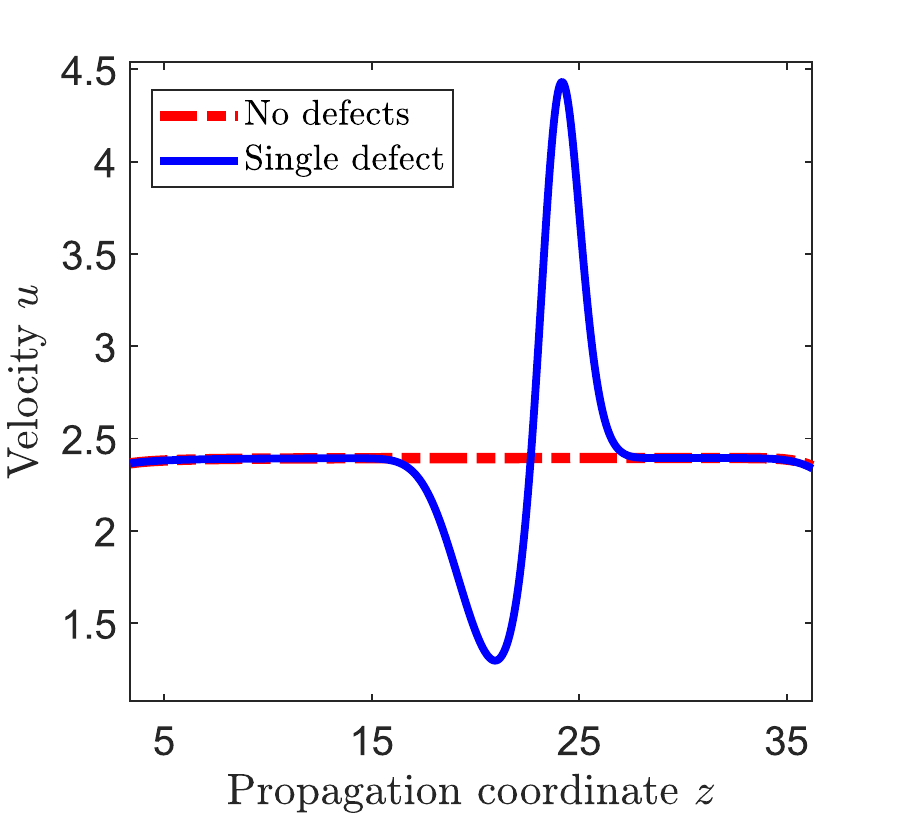}
	\centering
	\label{figure1}
	\caption{Comparison of the velocity \(u\) of wave propagation in two cases of a periodic Hatano-Nelson lattice  with \(N=100\) channels and \(h=0.6\): (1) no defects (red line), and (2) a single defect of strength $V=8$ located at site \(n=50\) (blue line).}
\end{figure}

In order to further gain insight regarding the underlying physical mechanism of this unexpected behavior, we compare the propagation dynamics of a periodic Hatano-Nelson lattice of \(N=100\) without  disorder (\(W=0\)) to that of a periodic Hatano-Nelson lattice with a single real defect, i.e., \(\epsilon_{n} = V\delta_{n,50}\), that induces a localized defect eigenstate.

As shown in Fig. 11, the presence of the defect leads to a deceleration of the wave as it reaches the defect site, followed by acceleration. However, the average velocity along the unidirectional propagation interval is greater for the case of an existent single defect. For example, for a lattice with \(h=0.6\), the velocity is \(u=2.39\) when the defect is absent. In fact, the average velocity increases to \(u=2.40\) for \(V=4\), and further increases to \(u=2.41\) when the defect strength is \(V=8\) (Fig. 11). Such a behavior provides a qualitative explanation of the increased average velocity for increasing disorder, where numerous localized eigenstates are typically found within the bulk of the lattice.

%%%%%%%%%%%%%%%%%%%%%%%%%%%%%%%%%%%%%%%%%%%
%%%%%%%%%%%%%%%%%%%%%%%%%%%%%%%%%%%%%%%%%%%

\section{Discussion and conclusions}
In conclusion, within the context of non-Hermitian photonics, we study the antagonism between two effects that occur in a Hatano-Nelson lattice with on-site diagonal real disorder. More specifically, on the one hand disorder induces localization and thus the wave tends to remain localized close to the excitation site, while on the other hand the asymmetric couplings favor propagation along the one edge of the lattice. We show that the interplay between these effects results in three qualitatively distinct phases of wavepacket's diffraction, namely localized, intermediate and delocalized, based on statistical averaging over many realizations of disorder. In addition, we find counter-intuitive characteristics regarding the relation between the strength of disorder and wavepacket's velocity, specifically increased velocity with increasing strength of disorder. Our results may provide useful insights regarding enhanced transport of waves in complex media, that is highly relevant to disordered photonics. 

%%%%%%%%%%%%%%%%%%%%%%%%%%%%%%%%%%%%%%%%%%%
%%%%%%%%%%%%%%%%%%%%%%%%%%%%%%%%%%%%%%%%%%%

\begin{acknowledgments}
Authors acknowledge financial support by the
European Research Council (ERC-Consolidator) under grant agreement No. 101045135 (Beyond Anderson). This research project was also co-funded by the Stavros Niarchos Foundation (SNF) and the Hellenic Foundation for Research and Innovation (H.F.R.I.) under the
5th Call of “Science and Society” Action Always strive for excellence – "Theodoros Papazoglou” (Project Number:11496,”PSEUDOTOPPOS”). Computations for this paper were partially conducted on the Metropolis cluster, supported by the Institute of Theoretical and Computational Physics at the Department of Physics, University of Crete. \end{acknowledgments}

%\nocite{*}

%\bibliography{apssamp}% Produces the %bibliography via BibTeX.

\end{document}